\documentclass[preprint]{emulateapj} 

\usepackage{natbib}
\usepackage{hyperref}
\hypersetup{colorlinks,citecolor=Blue,linkcolor=Red,urlcolor=Blue}
\usepackage{amsmath}
\usepackage{empheq}
\usepackage[usenames,dvipsnames]{color}
\allowdisplaybreaks 

\newcommand{\icarus}{Icarus}

\begin{document}
 
\title{Observational constraints on the orbit and location of Planet Nine in the
outer solar system}
\author{Michael E. Brown \& Konstantin Batygin} 

\affil{Division of Geological and Planetary Sciences, California Institute of Technology, Pasadena, CA 91125} 
\email{mbrown@caltech.edu, kbatygin@gps.caltech.edu}
 
\newcommand{\Ham}{\mathcal{H}}
\newcommand{\G}{\mathcal{G}}
\newcommand{\appropto}{\mathrel{\vcenter{\offinterlineskip\halign{\hfil$##$\cr\propto\cr\noalign{\kern2pt}\sim\cr\noalign{\kern-2pt}}}}}

\begin{abstract} 
We use an extensive suite of numerical simulations to constrain the 
mass and orbit of Planet Nine, the recently proposed perturber in a distant
eccentric orbit in the outer solar system. We 
compare our simulations
to the observed population of aligned eccentric high semimajor axis Kuiper
belt objects and determine which simulation parameters are statistically
compatible with the observations. We find that only a narrow range
of orbital elements can reproduce the observations. In particular,
the combination of semimajor axis, eccentricity, and mass
of Planet Nine
strongly dictates the semimajor axis range of the orbital 
confinement of the distant eccentric Kuiper belt objects. 
Allowed orbits, which confine Kuiper belt objects with semimajor axis
beyond 380~AU, have perihelia roughly between 150 and 350~AU,
semimajor axes between 380 and 980~AU, and masses between 5 and 20 Earth masses.
Orbitally confined objects also generally have orbital planes
similar to that of the planet, suggesting that the planet
is inclined approximately 30~degrees to the ecliptic. 
We compare the allowed orbital positions and 
estimated brightness of Planet Nine to previous and ongoing surveys
which would be sensitive to the planet's detection and use
these surveys to rule out approximately two-thirds of the planet's orbit.
Planet Nine is likely near aphelion with an approximate brightness of
$22<V<25$. 
At opposition, its motion, mainly due to parallax,
can easily be detected within 24 hours.
\end{abstract}

\section{Introduction}
Since the time of the discovery of Sedna, it
has been clear that a large perturbing mass either is or was present in
the outer solar system at some time \citep{2004ApJ...617..645B}. 
With a perihelion distance of 76~AU, Sedna is essentially 
immune to direct interactions with the known planets, thus, 
unlike all other Kuiper belt object orbits, 
it cannot have been placed onto its orbit by perturbation from
any of the known planets.
Proposals for the perturber required to have created Sedna's orbit 
 have included sibling stars in the sun's birth
cluster \citep{2004ApJ...617..645B,2006Icar..184...59B,2012ApJ...754...56D}, 
a single passing star \citep{2004AJ....128.2564M, 2004Natur.432..598K,
2004A&A...428..673R}
as well as a small former or extant planet  in the outer solar system
\citep{2004ApJ...617..645B, 2006ApJ...643L.135G, 2006Icar..184..589G}.
Progress on understanding the cause of Sedna's perturbed 
orbit, however, was not possible because of a lack of 
additional high perihelion objects.

With the discovery of 2010 GB174 \citep{2013ApJ...775L...8C} 
and 2012 VP113
\citep{2014Natur.507..471T}
 -- the second and third high perihelion Sedna-like objects 
-- additional patterns began to emerge. 
Most importantly, \citet{2016AJ....151...22B} 
point out that all well-determined orbits
of Kuiper belt objects (KBOs) beyond Neptune
with semimajor axis, $a$, larger than 227~AU approach perihelion
within 94~degrees of longitude of each other.
Moreover, these objects also share
 very nearly the same orbital plane, which is tilted 
an average of 22~degrees to the ecliptic. The combined probability 
of these two occurrences happening simply due to chance is less than
0.01\%. Importantly, of all KBOs with $a>100$~AU, the five with the
largest perihelion distances are likewise confined to the same 
perihelion region
and orbital plane.

\citet{2016AJ....151...22B} show that
a distant massive eccentric planet will cause clustering of the
perihelion and orbital planes of distant Kuiper belt objects
in the manner observed, and, additionally, will naturally lead
to the creation of objects with high perihelion
orbits like Sedna. 
Surprisingly, these clustered and high perihelion objects
have orbits that are anti-aligned with the giant planet. That
is, the clustered Kuiper belt objects come to perihelion 180~degrees
away from the perihelion position of the planet. Despite chaotic evolution, the crossing orbits
maintain long term stability by residing on a interconnected web of phase-protected mean motion resonances.

The distant eccentric perturber studied in \citet{2016AJ....151...22B} -- which 
we refer to as Planet Nine -- modulates the perihelia of objects
in the anti-aligned cluster and naturally creates objects like 
Sedna, in addition to the other high perihelion KBOs. 
Additionally, the existence of Planet Nine
predicts a collection of high semimajor axis eccentric objects with 
inclinations essentially perpendicular to the rest of the solar system.
Unexpectedly, this prediction is strongly supported by the 
collection of low perihelion Centaurs with perpendicular orbits whose
origin had previously been mysterious \citep{2015Icar..258...37G}. 

Here we make detailed comparisons between dynamical simulations 
that include the effects of Planet Nine,
and solar system observations, to place constraints on the orbit
and mass of the distant planetary perturber. We then discuss observational
constraints on the detection of this distant giant planet and future 
prospects for its discovery. 

\section{Constraints on mass, semimajor axis, and eccentricity}

The inclined orbits of the aligned KBOs 
(and thus, presumably, of the
distant planet) render ecliptic-referenced 
orbital angles awkward to work in (particularly when we consider the
Centaurs with perpendicular orbits later). Accordingly, we re-cast the three
ecliptic-referenced parameters -- argument of perihelion, longitude of ascending
node, and inclination -- into simple descriptions of orbit in absolute
position on the sky: the ecliptic longitude of the point in the sky 
where the object is at perihelion (which we call the ``perihelion longitude'', 
not to be confused with the standard 
orbital parameter called ``longitude of perihelion'' which, confusingly,
 does not actually measure the longitude of the perihelion except for zero inclination orbits), the latitude of the perihelion (``perihelion
latitude''),
and an angle which measures the projection of the orbit pole onto the
plane of the sky (``pole angle'' perhaps more easily pictured
as the direction
perpendicular to the motion of the object at perihelion). 

Figure 1a shows the perihelion longitude and latitude as well as the pole
angle for all objects with $q>30$ and $a>60$~AU and well-determined orbits. The seven
objects with $a>227$~AU are highlighted in red.
The clustering in perihelion location as well as pole angle is clearly visible.
In Figure 1b 
we plot the perihelion longitude of all well constrained orbits in the Kuiper 
belt which have perihelia beyond the orbit of Neptune as a function of
semimajor axis. The 7 objects with $a>227$~AU cluster within 94~degrees
of perihelion longitude.
\cite{2016AJ....151...22B} showed that this clustering,
when combined with the clustering in pole angle, was unexpected at the 99.993\%
confidence level. The clustering is
consistent with that expected from a giant planet whose perihelion  is located
180~degrees in longitude away from the cluster, or an ecliptic
longitude of $241 \pm 15$~degrees. A closer examination of Figure 1b makes clear
a possible additional unlikely phenomenon. While KBOs with semimajor axes out
to 100~AU appear randomly distributed in longitude, from 100 to 200~AU, 13
objects are loosely clustered within 223~degrees of each other. While this
clustering is not as visually striking, the probability of such a loose
clustering of 13 objects occurring in randomly distributed data is smaller than 5\%.
As will be seen, such a loose clustering of smaller semimajor axis objects
can indeed be explained as a consequence of some orbital 
configurations of a Planet Nine.

\begin{figure}
\plotone{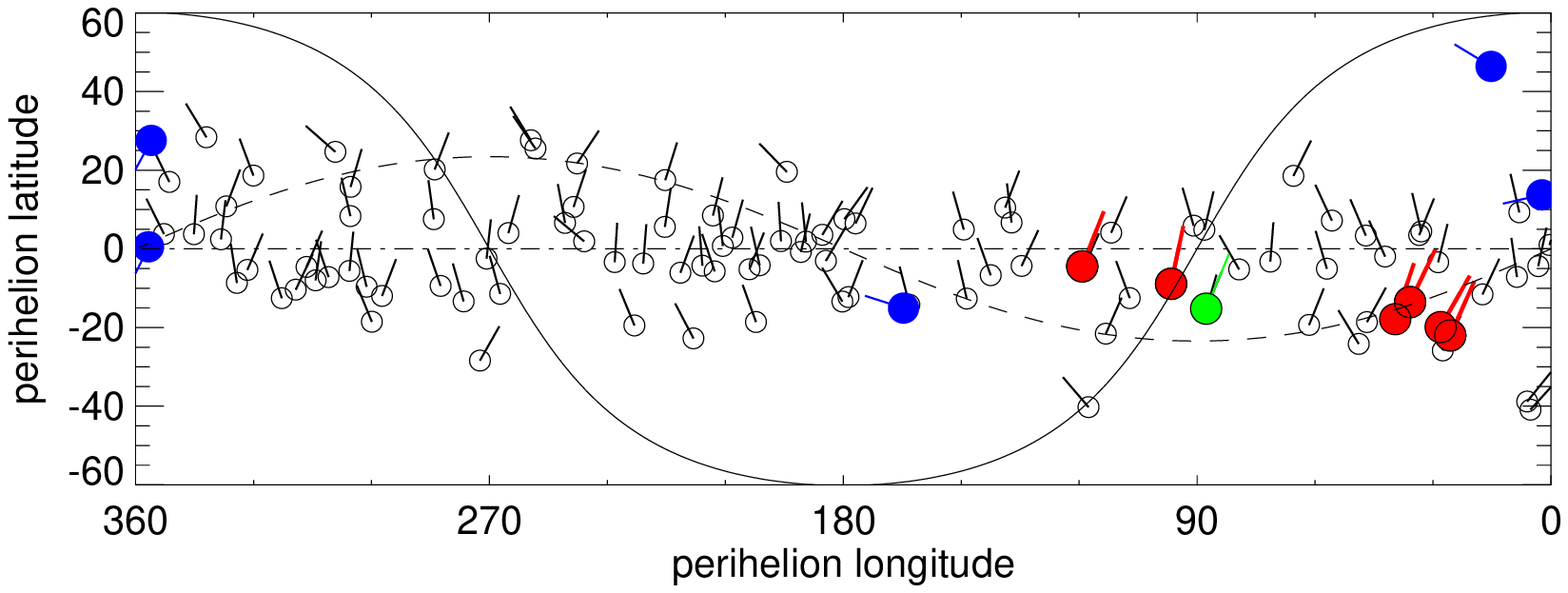}
\plotone{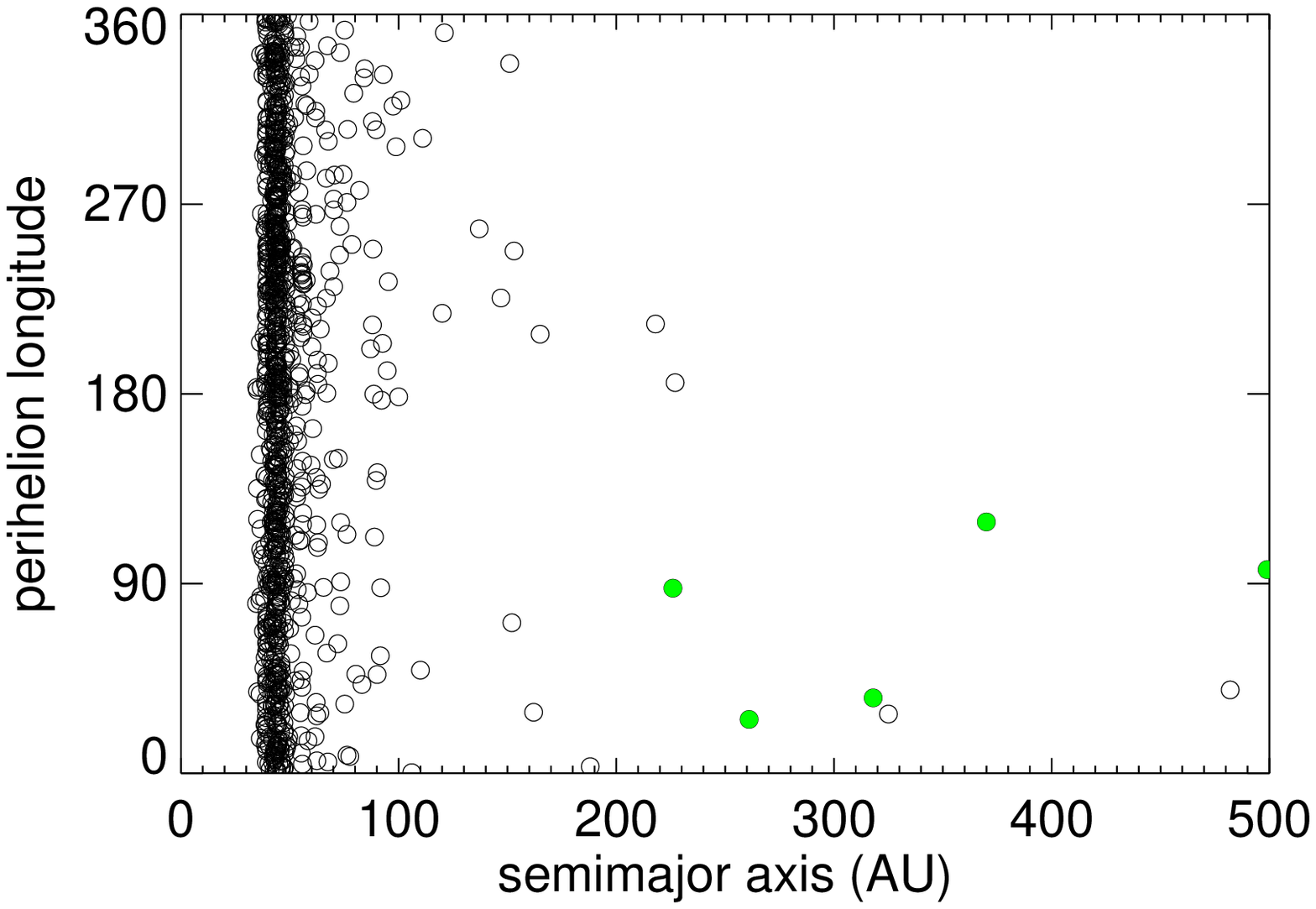}
\caption{Orbital parameters of distant Kuiper belt objects. (a) 
The standard orbital parameters
argument of perihelion, longitude of ascending node,
and inclination can be transformed into non-standard, but more 
readily interpretable ecliptic longitude and latitude of the
point where the object comes to perihelion and an angle which
is a projection of the orbits pole position on the sky. In this
representation, the collection of all objects with $q>30$ and $a>60$~AU
is shown. The six objects with the highest semimajor axis are highlighted in
red. The KBO 2000 CR105, which has the seventh largest semimajor axis
and has an elevated perihelion of 44~AU, is shown in green. The blue 
points are all of the object with $a>227$~AU and $i>50$~degrees. All
of these objects are Centaurs with perihelia inside of 15~AU.
(b) A plot of the semimajor axis versus the ecliptic 
longitude at which the object comes to perihelion for all
KBOs with well determined orbits and with
$q>30$~AU shows that the seven KBOs with the largest semimajor
axes are clustered within 94~degrees of each other. The
green points additionally highlight all objects with $a>100$~AU and $q>42$~AU,
showing that these objects, too, are similarly clustered.
The weaker potential clustering of objects $\sim$180~degrees away
is also evident between 100 and 200~AU.}
\end{figure}

To understand how these  observations constrain the mass and orbit of
Planet Nine, we performed a suite of 
evolutionary numerical integrations. Specifically, we initialized a 
planar, axisymmetric disk consisting of 400 eccentric planetesimals, 
that uniformly spanned semimajor axis and perihelion, $q$, distance ranges of 
$a=150-550$~AU and $q=30-50$~AU, respectively.
The planetesimal population (treated as test particles) was evolved 
for 4~Gyr under the gravitational influence of the known giant planets, 
as well as Planet Nine.

Perturbations due to Planet Nine and Neptune were accounted for in a 
direct N-body fashion, while the secular effects of the remaining 
giant planets were modeled as a suitably enhanced quadrupolar 
field of the Sun. As shown in \citet{2016AJ....151...22B}, such a numerical 
setup successfully captures the relevant dynamical phenomena, 
at a substantially reduced computational cost. 

In these
integrations we varied the semimajor axis and eccentricity of Planet Nine from
$a_9=200-2000$~AU and $e_9=0.1-0.9$ in increments of $\Delta a_9=100 $~AU and 
$\Delta e_9=0.1$ (here, and subsequently, the 9 subscript
refers to the orbital parameters of Planet Nine, while unsubscripted 
orbital parameters refer to the test particles). 
The $a_9-e_9$ grids of synthetic scattered disks were constructed 
for Planet Nine masses of $m_9=0.1$, $1$, $10$, $20$ and $30$ $M_e$ (Earth masses), 
totaling a suite 
of 320 simulated systems. All calculations were performed using the 
\texttt{mercury6} N-body integration software package \citep{1999MNRAS.304..793C}, 
employing the hybrid symplectic-Bulirsch-Stoer algorithm with a timestep 
equal to a tenth of Neptune's orbital period.

We assess the success of each simulation with two simple metrics.
First,
we collect the orbital elements of all remnant objects at each 0.1 Myr output
time step 
from 3 to 4~Gyr after the
start of the simulation (in order to assure that the objects we are considering
are stable over at least most of the age of the solar system), and
we restrict ourselves to objects with instantaneous
perihelion $q< 80$~AU (to restrict
ourselves to objects which are most likely to be observable). In this 
fashion we are examining stream functions of orbital elements which fit into
an observable range of parameter space, rather than examining individual
objects at a single time step. This approach is used in all subsequent 
discussions of simulations below. We then 
select 13 objects at random in the $a=[100,200]$~AU 
range and 7 objects at
random in the $a=[227,600]$~AU range and calculate the 
smallest angles that can be used to confine the two populations. We
perform this random selection 1000 times and calculate the joint probability
that, like the real data, the 13 objects in $a=[100,200] $AU range are confined within 223~degrees
and the 7 objects $a=[227 - 600]$~AU range are confined within 94~degrees.
Additionally, we examine whether {\it any} objects exist in the 
range $a=[200,300]$~AU, as many simulations remove all objects in this range.
We assign these simulations
a probability of zero. This probability calculation has
the advantage that it is agnostic as to whether or not our observations
of clustering are
significant or even physically relevant.
It simply calculates the probability that a given 
simulation could reproduce some of the apparent features of the real data,
even if by chance.

The second metric we use to assess the success of the simulations 
relies on the observation of \citet{2016AJ....151...22B} that a distant
massive eccentric perturber will cause secular perturbations which
lower the eccentricity and thus raise the perihelion of moderate
semimajor axis objects at a wide range of perihelion latitudes.
This effect increases strongly with increasing perturber
eccentricity and
with decreasing perturber semimajor axis. Of the 15 known KBOs
with $100<a<220$ AU and $q>30$, zero have perihelion greater than 42 AU
(whereas 5 of the 7 objects with $a>227$ AU have such elevated perihelia).
These simulations were not designed in such a way that simple assessment
of the probability of such a distrubtion is straightforward. 
We approximate this probability by collecting all objects from time steps
between 2 and 4 Gyr with $30<q<40$ AU and $100<a<200$ AU, randomly
selecting 15 objects from this set, and calculating the product
of the fraction of the time that each of these objects spends with $q<42$ and
$100<a<200$~AU. These probabilities range from unity, when Planet Nine
is distant or only mildly eccentric, to $10^{-12}$ for higher eccentricities
and lower semimajor axes. Because these calculated probabilities are
difficult to straightwordly compare to the real data, we instead
impose a simple threshold and assert that simulations with a 99\% or higher
probability of creating at least one
high perihelion object in the $100<a<200$ AU
range are effectively ruled out. We assign their overall probability
to zero. 

Figures 2 and 3 show grids of probabilities for the 10 and 20 $M_e$ simulations 
(the 0.1, 1, and 30 $M_e$
simulations have no acceptable solutions). 
The probabilities should be taken as more
qualitative than quantitative, as these simulations are exploratory and
attempt to cover large ranges of phase space by including
a limited number of particles per simulation
and excluding three dimensional effects. Nonetheless, the overall 
trends are clear. The lack of high perihelion objects between
$100<a<200$ AU strongly rules out all of the low semimajor axes and
nearly all of the highest eccentricities, while the need to confine
objects in perihelion longitude requires moderately high eccentricities
or moderately low semimajor axes. The combined effect of
these two constraints makes for a rather narrow combination
of $a_9$ vs. $e_9$ with acceptable solutions. In fact, the range is
sufficiently narrow that the it is clear that the grid spacing of 
our simulations is often too large to capture acceptable solutions
at all semimajor axes or eccentricities.

\begin{figure}
\plotone{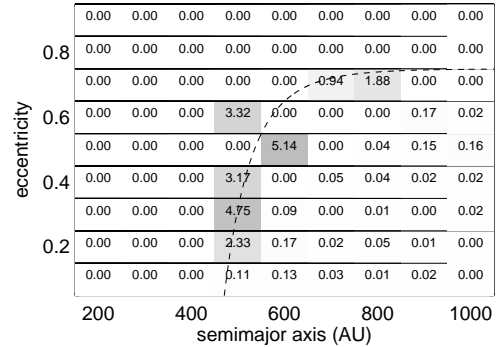}
\caption{Only a limited set of simulated 10~M$_e$ Planet Nine orbits provide an adequate fit to the known orbits of the distant KBOs. For each semimajor axis - 
eccentricity combination, we calculate the probability that 7 objects
selected at random with $227<a<600$, $q<80$~AU and at times
$t>3$~Gyr 
are clustered within
94~degrees of each other combined with the probability that 13 objects
randomly selected with $100<a<200$ and $q<80$~AU. In addition we
discard simulations with an unacceptably large probability of creating
high perihelia in low semimajor axis objects.
Acceptable simulations
(with a probability of greater than 1\%) occupy a narrow range
of $a_9$-$e_9$ space.}
\end{figure}

\begin{figure}
\plotone{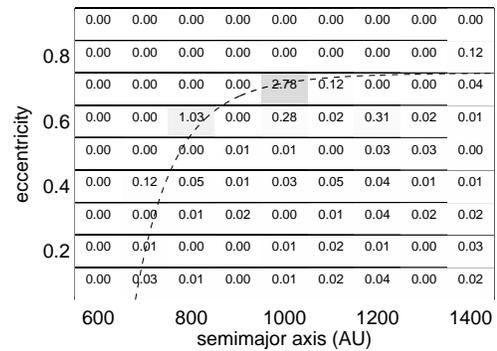}
\caption{For a 20 M$_e$ Planet Nine with $a_9=[800,1000]$~AU 
and $e_9=[0.6,0.7]$ a marginal
fit appears. At higher masses no simulations can reproduce the
observations.}
\end{figure}

Nonetheless, from the 10 $M_e$ simulations we can discern 
the narrow range of acceptable results 
(Figure 2).
In general, all simulations which cannot be excluded at at the 99\%
confidence level fall 
with a range of $a_9=[500, 800]$~AU approximately along an empircally
defined line of $e_9=0.75-(450 {\rm AU}/a_9)^8$, shown as the dashed line
in Figure 2. Such orbits have a perihelion,
$q_9$, in the range $\sim [200,340]$~AU.
The acceptable $q_9$ range is greater at smaller semimajor axis; at 600 AU
and beyond, all solutions have $q_9 \sim 200$ AU.

For the 20 $M_e$ simulations, the locus of acceptable $a$-$e$ combinations
shifts outward and can be fit with a similar empircal function
 $e_9=0.75-(650 {\rm AU}/a_9)^8$, but
only from $a_9=[800,1000]$ AU. Lower semimajor axes perturb the
low semimajor axis KBOs to high perihelia, while higher semimajor axes
fail to cluster the high semimajor axis KBOs appropriately.
Higher mass simulations cannot 
match the observations at all.

In Figure 4 we show, as an example, the Planet Nine-centered
 perihelion longitudes as a function of $a$ of
all objects that have $q<80$~AU and which have survived at least 3~Gyr,
for the case of $M_9=10 M_e$, $a_9=500$~AU, and $e_9=0.6$, one of the 
simulations along the acceptable locus. 
Objects anti-aligned with the planet have a longitude
of 180~degrees in this simulation, while those aligned will be at 0~degrees
longitude. This simulation shows the major effects
that we have previously identified in the real data. Inside of 100~AU
little perturbation is visible. From 200 to 600~AU the longitudes are
confined around 180~degrees, that is, they are anti-aligned with Planet Nine.
And from 100 to 200~AU there is a slight tendency for a broad cluster
centered on the longitude of the planet.
\begin{figure}
\plotone{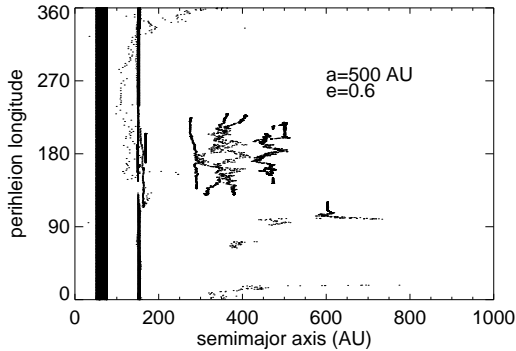}
\caption{One of the successful simulations showing the orbital evolution of
all objects with $q<80$~AU at times $t>3~Gyr$.
A collection of resonant stable high semimajor axis objects 
anti-aligned with Planet Nine (centered at $\Delta$ longitude of
180~degrees) appears beyond 200~AU, while from 100-200~AU there is a slight
preference for aligned orbits.}
\end{figure}

As a counter example, Figure 5 shows a simulation with $M_9=10 M_e$, $a_9=700$~AU and 
$e_9=0.3$
which depicts many of the same general phenomena, but these phenomena do not
develop until larger semimajor axes. For example, the anti-alignment does not
begin until 400~AU, while a broad aligned 
cluster can be seen
from about 300 to 400~AU. Even with the crudeness of these simulations
these basic effects are clear. The semimajor axis at which anti-alignment
begins and the range where broad confinement is evident are strong
indicators of the combination of 
semimajor axis and the eccentricity of Planet Nine.
\begin{figure}
\plotone{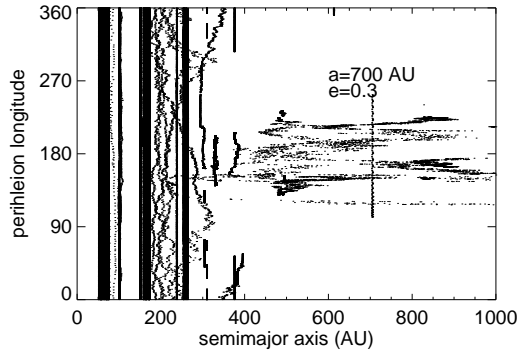}
\caption{A low probability simulation. Clear anti-alignment does not 
develop until beyond 400~AU, while aligned orbits appear only from
approximately 300 to 400~AU.}
\end{figure}

\section{Constraints on inclination and argument of perihelion}
The planar simulations provide no constraints on inclination, $i_9$,
argument of perihelion $\omega_9$, or longitude of ascending 
node, $\Omega_9$, of Planet Nine. 
To examine the effects of these orbital
elements on the Kuiper belt, we perform a second, fully three dimensional
suite of simulations. In these simulations we fix the semimajor
axis and eccentricity to be 700~AU and 0.6, respectively, values
which are within our acceptable range of parameter space. The 
inclination dynamics are unlikely to be unique to the
specific values of $a_9$ and $e_9$, so we deem these simulations
to be representative. We allow the inclination of Planet Nine to
take values of $i_9=$ 1, 10, 20, 30, 60, 90, 120, and 150~degrees.

Unlike the planar suite of calculations, here we 
start all planetesimals with their longitude of perihelia anti-aligned
to that of Planet Nine.
The starting values of planetesimals' longitudes of ascending node, 
on the other hand, are taken to be random. 
As demonstrated by \citet{2016AJ....151...22B}, 
dynamical sculpting of such a planetesimal population 
yields a configuration where long-term stable objects 
have longitudes of ascending node roughly equal to that of Planet Nine. 
In turn, this ties together $\omega_9$ and $\Omega_9$ 
through a fixed longitude of perihelion (which is the sum of these parameters). 

Upon examination of these simulations, we find that efficiency of
confinement of
the distant population drops dramatically with increased inclination
of Planet Nine. 
To quantify this efficiency, we sample each simulation 1000 times, picking
7 random objects from the sample of all objects in the range $a=[300,700]$~AU
(as previously shown, these $a_9=700$~AU simulations do not begin strong
perihelion confinement until $a\sim$300~AU; as we are more interested in
understanding the cluster than in specifically simulating our data at
this point we increase our semimajor axis range of interest).
As before we restrict ourselves to time steps after 3~Gyr in which an object's orbit
elements have $q<80$~AU, and we add the constraint that $i<50$~degrees,
to again account for observability biases.
We calculate the fraction of times that the 7 randomly selected
 objects are
clustered within 94~degrees (Figure 6). The confinement
efficiency drops smoothly until, at an inclination of 60~degrees
and higher, it scatters around 20\%. These results suggest, but do not
demand, that Planet Nine has only a modest inclination.

\begin{figure}
\plotone{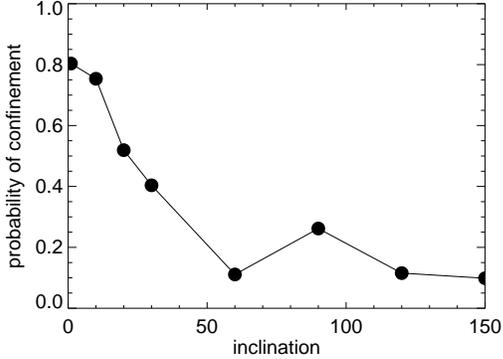}
\caption{For simulations with $a_9=700 $AU and $e_9=.6$, 
the probability of confinement of 7 randomly selected objects
with $300<a<700$ and $q<80$~AU varies strongly with inclination of the planet.}
\end{figure}

One of the striking characteristics of the 7 aligned distant Kuiper
belt objects is the large value of and tight confinement in the pole angle (22 $\pm$ 6~degrees; Figure 1). 
In examining the simulations that
exhibit good confinement in longitude, we note that the polar angles
of the simulated orbits are
approximately perpendicular to the plane of Planet Nine, particularly 
for objects which come to perihelion in the plane of the planet. 
The implication of this phenomenon is that a pole angle of 22~degrees
suggests a minimum planet inclination of approximately 22~degrees,
and an orbital plane (which is controlled by $\Omega_9$)
similar to the plane of the observed objects.
To quantify this observation, we plot the median pole angle of our
simulation objects which meet the criteria described above and which
have perihelion latitudes between -25 and 0~degrees like the real distant 
objects.
We restrict ourselves to objects with perihelia south of the ecliptic
both because the observed objects all have perihelia south of the
ecliptic and also because we want to avoid any bias that would occur
by a loss of observed objects north of the ecliptic due to the proximity
of the galactic plane close to the perihelion positions. It is currently
unclear whether the lack of clustered objects with perihelia north of
the ecliptic is a dynamical effect or an observational bias.
Figure 7 shows this mean polar angle as a function of 
argument of perihelion of Planet Nine. 
The maximum median polar angle occurs for
$\omega_9\sim 150$~degrees, a configuration where the plane of the planet 
passes through the perihelion positions of the objects, 
and that maximum is approximately
equal to the inclination of
the orbit, confirming our observation that the clustered objects are
along the same orbit plane as the planet.

\begin{figure}
\plotone{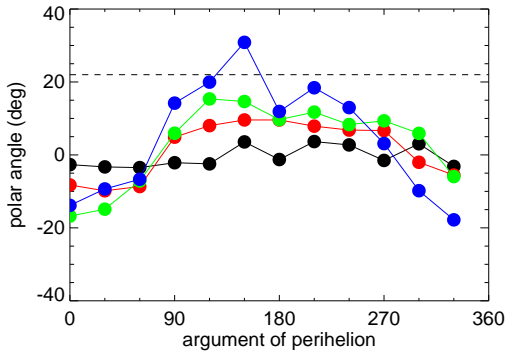}
\caption{For the inclined simulations, the average polar angle of anti-aligned
orbits varies systematically with $\omega_9$.
The average
polar angle of the seven most distant objects is 22~degrees, shown as the dashed
line. For simulations with high probability of confinement ($i_9<60$~degrees),
the average polar angle only reaches values as high as 22~degrees for
$\i_9=30 $~degrees and $\omega_9$=150~degrees. The black, red, green, and
blue points correspond to the $i_9$=1, 10, 20, and 30 degree simulations, respectively.}
\end{figure}

Based on the confinement probability and the large average pole angle of
the real objects, we can infer that the inclination of Planet Nine is 
greater than $\sim$22~degrees and less than the inclination at which confinement becomes improbable, which, based on an interpolation of the data from 
Figure 5, occurs approximately around 40~degrees. For inclinations
of $\sim$22~degrees, $\omega$ must be quite close to 150~degrees.
For inclinations of 30~degrees, the allowable range for
the argument of
perihelion appears to be  $\omega_9 \sim 120-160$~degrees.

While this analysis yields useful constraints, we quantify 
these results further by again sampling
each of the simulations 1000 times and determining the probability
of 6 randomly selected objects (300~AU $<a<$ 700~AU, $q< 80$~AU, 
$i<50$
degrees, survival time greater than 3~Gyr, and perihelion latitude between
-25 and 0~degrees) having perihelion longitudes clustered with 94~degrees
and having an average polar angle greater than 20~degrees with an
RMS spread of less than 6.2~degrees. Almost all simulations can be ruled
out at greater than the 99\% confidence level. The only simulations
which cannot are, unsurprisingly, those with inclination of 30~degrees
and argument of perihelion of 150 -- which is the single best fit -- and
120~degrees and, additionally, a few other seemingly
random combinations of [$i_9$, $\omega_9$]: [90, 60],
[150, 0], [150, 210], and [150, 330], all in units of degrees. 
We examine all of these cases in detail below.

One strong prediction of the existence of a giant planet in the outer
solar system is that it will cause Kozai-Lidov oscillations which will
drive modest inclination objects onto high inclination perpendicular and
even retrograde orbits and then back again. This effect can be
seen, for example, in Figure 8, where we plot the evolution of
perihelion longitude
versus inclination for the simulations with 30 degree inclination
 (again, restricting
ourselves to 300$>a>$700~AU, $q<80$ AU and $t>3$~Gyr; note that argument of
perihelion of Planet Nine has no substantive effect on this plot, so we plot all arguments
together). The five known objects
in the outer solar system with $a>200$~AU and $i>50$ deg are also shown.
The simulations reproduce their perihelion longitudes and inclinations well,
although they are all on the outer edge of the predicted clustering regions.
An important caveat to note, however, is that the five high inclination
objects are all Centaurs with $8<q<15$~AU. The high inclinations of these
objects mean that they penetrate the giant planet region much more 
easily and so can maintain their alignments much more easily than lower
inclination Centaurs. Our simulations remove all objects inside 20~AU,
so we have not explored the dynamics interior to Uranus, but we note a
systematic trend where objects with smaller perihelion distances move 
to the outer edge of the clustering regions, just like the real 
low perihelion objects appear to be. Clearly, simulations including all
of the giant planets which allow us to study these high semimajor
axis Centaurs are critical.
\begin{figure}
\plotone{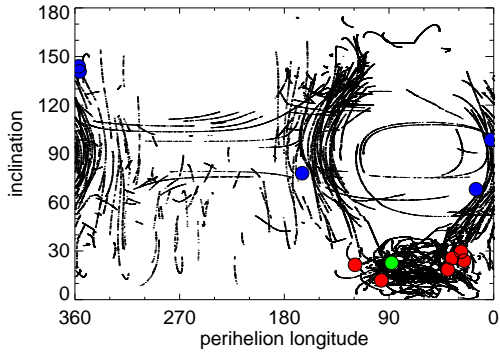}
\caption{For simulations with $a_9=700 $~AU, $e_9=0.6$, and $i_9=30 $ deg, 
examination of all objects with $300<a<700$~AU and $q<80$ AU shows
a low inclination population anti-aligned with Planet Nine and a high
inclination population on either side of the anti-aligned population.
The colored points are the same as in Figure 1a, with the blue points
showing that the highly inclined large semimajor axis Centaurs are closely
aligned with the predicted high inclination locations.}
\end{figure}

The perihelion locations of the perpendicular high semimajor axis Centaurs
effectively rule out the possible higher inclination orbits for Planet Nine.
The $i_9=90$ and $i_9=150$~degrees cases do create high inclination objects, 
but their perihelia are sporadically distributed across the sky (Figure 9). We conclude
that, of our simulated parameters, only the $i_9=30, \omega_9=150$~degrees and 
$i_9=30, \omega_9=120$~degrees are viable.
\begin{figure}
\plotone{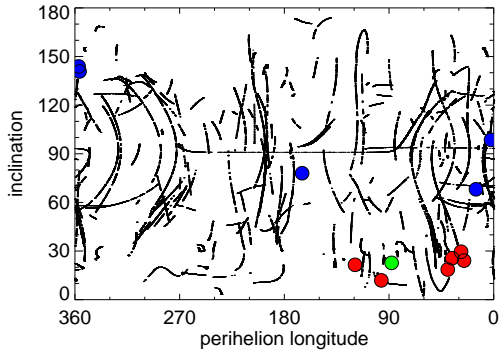}
\caption{
The simulations with $i_9=150$~degrees produce little confinement of
either the low or high inclination population.}
\end{figure}

In order to complete our analysis in a tractable amount of time, 
each of the simulations used to explore parameter space above was limited
in either dimensionality, number of particles, or in the range of
starting parameter of the particles. In order to check that these limitations
did not influence the overall results, we perform a final fully three
dimensional simulation with a large number of particles with randomly chosen
starting angles. We choose to simulate Planet Nine with a mass of 
10~M$_e$, $a_9=700$, $e_9=0.6$, $i_9=30$~degrees, 
$\omega_9=0$~degrees, and $\Omega_9=0$~degrees (note that the planet precesses over 
the 4 billion years of the integrations, but at this large semimajor axis
the precession in $\omega_9$ is only about 30~degrees during the entire period,
so we ignore this effect). These full simulations reproduce all of the relevant
effects of the more limited simulations, giving confidence 
to our simulation results.

\section{Sky position}
Based on comparison to our suite of simulations,
we estimate that the orbital elements of Planet Nine are as follows.
For 10 and 20 M$_e$ planets, $a_9$ and $e_9$ are empirically defined
in Section 2. We roughly bound these empirical functions with simple
linear fits as a function of mass
on the minimum and maximum semimajor axes of Planet Nine
for the two masses and a simple linear fit to the empirical fitting function.
We thus estimate that $a_9$ is in the range [200AU+30$M_9/M_e$,600AU+20$M_9/M_e$],
where $M_9$ is approximately in the range [5 $M_e$, 20 $M_e$],
and that $e_9$=0.75-$[(250{\rm AU}+20M_9/M_e$)/$a_9]^8$

The inclination is between approximately 
$22<i_9<40$~degrees, and the argument of perihelion is 
between $120<\omega_9<160$~degrees. We fix the perihelion longitude
at 241$\pm$15~degrees.
While these choices of parameter ranges have been justified in the 
analysis of the simulations above, they cannot be considered a 
statistically rigorous exploration of parameter space.
Indeed, any attempt at such statistical rigor is not yet warranted:
substantial uncertainty comes not just from the statistics of the objects
themselves, but from the currently
small number of simulations in the best fit region of parameter space. 
Clearly, significantly more simulation is critical for a better
assessment of the path of Planet Nine across the sky.

The last parameter we consider is the mass of Planet Nine, which 
we assume is in the range of 5 to 20 M$_e$. To transform this mass into
an expected brightness requires assumptions of both radius (and thus 
composition) and albedo (and thus surface composition), neither of
which is constrained by any of our observations. 
\citet{2016arXiv160407424F} consider a range of Planet Nine masses and core fractions.
For masses between 5 and 15 M$_e$ with 10\% atmospheric mass fraction,
the radius is approximately fit as $R_9=[2.4+0.1 (M_9/M_e)]$ $R_e$, which 
we will use as our nominal relationship through this range and extending
to 20 M$_e$. In addition
\citet{2016arXiv160407424F} find a quite high albedo value of 0.75 primarily due to
Rayleigh scattering in the atmosphere. We assume, to be conservative, a 
range between 0.3, the approximate albedo of Neptune, and the modeled value
of 0.75.

We now use our estimated orbital parameters to predict the
orbital path of Planet Nine across the sky. We carry out a simple Monte Carlo
analysis selecting uniformly across all of the parameter ranges.
Figure 10 shows the sky location, heliocentric distance, magnitude,
and sky motion
at opposition for our suite of predicted orbits. 

\begin{figure}
\plotone{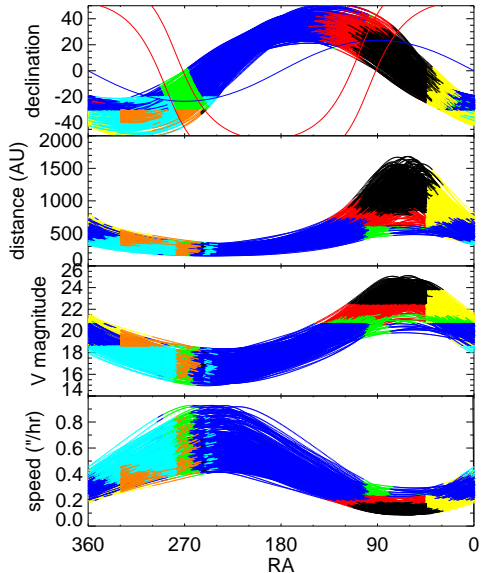}
\caption{Using all constraints on the orbital and physical
parameters of Planet Nine, we 
can predict the location, distance, brightness, and speed
of the planet throughout its orbit. Regions within 10~degrees of
the galactic plane are outlined in red, and the ecliptic plane is
shown in blue. The colored portions show regions where Planet Nine 
would have been or should be detected by previous or ongoing surveys.
Light blue shows limits from the CRTS reanalysis,
yellow shows Dark Energy Survey limits and coverage, dark blue shows 
Pan-STARRS transient analysis limits, green shows Pan-STARRS moving object
analysis current limits, and red shows eventual Pan-STARRS expected limits.
Orange shows the region exclusively ruled out by lack of 
observed perturbation to Saturn 
\citep{2016A&A...587L...8F,2016arXiv160403180H}.
The black regions show regions of phase space where Planet Nine could not have
been or will not be detected in previous or currently planned surveys.}
\end{figure}

\section{Current observational constraints}
While most wide field surveys of the Kuiper belt have not been
sensitive to sky motions
smaller than about 1 arcsecond per hour \citep[][i.e.]{2002AJ....123.2083M,
2008ssbn.book..335B,2011AJ....142..131P}, a
few surveys have had the sensitivity and cadence to have potentially
detected Planet Nine at some point in its orbit. We discuss all
such surveys below.

\subsection{WISE}
The WISE spacecraft surveyed the entire sky twice in its 3.4 and 4.6 
$\micron$ (W1 and W2)
bands, allowing \citet{2014ApJ...781....4L} to rule out Saturn-sized 
planets -- which have substantial enhanced short wavelength emission 
owing to emission from internal heat --
out to a distance of $\sim$30,000~AU. A 10~M$_e$ planet, however, would
not be expected to have this enhanced short wavelength emission. For example,
ISO detected nothing but reflected sunlight from Neptune 
from 2.5 to 5 $\mu$m \citep{2003Icar..164..244B},
with an average flux of about 5 mJy. The \citet{2014ApJ...781....4L} W1 limit 
corresponds to approximately 0.2 mJy, which suggests that Neptune
itself could only be detected to $\sim$70~AU. As a confirmation,
we examined the catalog of
WISE single-source detections of Neptune itself. Neptune is detected
16 times with a signal-to-noise (S/N) of approximately 50 in the W1 band.
Assuming that all of those images are coadded, Neptune could be detected
with a S/N of 10 only to $\sim$63~AU, consistent with our 
estimate above. 
\citet{2016arXiv160407424F} suggest that the W1 brightness of Planet Nine could be
enhanced above its blackbody level and thus potentially observable to a greater
distance, but the sensitivity of the WISE data to Neptune-sized planets except at
the closest possible distances remains low. 
The \citet{2014ApJ...781....4L} results thus provide 
no constraints on the position or existence of Planet Nine.

\subsection{Catalina real time transient survey}
While near earth object searches 
are performed at cadences poorly matched to the 
detection of 
objects in the outer solar system,
they
cover the sky multiple times in a year, allowing the possibility of
detecting objects by their weekly or monthly motion.
In \citet{2015AJ....149...69B}, we performed such an analysis from the Catalina real time 
transient survey \citep[][CRTS]{2009ApJ...696..870D}, 
which itself repurposed the Catalina Sky Survey
near earth asteroid search into a transient survey.  
We collected all one-time transients over an 8 year period, that is, all 
instances in which an object was detected at a spot in the sky only once,
and attempted to fit all combinations of 4 or more detections to Keplerian
orbits. The Keplerian filter is strong. From $\sim10^{19}$ 
potential combinations,
we narrowed the detections down to eight known Kuiper belt objects and
zero false positives. Every bright Kuiper belt object in the survey
fields was detected, often dozens of times. The survey was 
determined to be essentially 100\% complete to $V\sim 19.1$ in the north
and $V\sim 18.6 $ in the south. For some of the smaller potential values
for semimajor axis and larger values of planetary radius, 
for example, Planet Nine would have been 
visible to this survey over a substantial portion of its orbit, 
though it was not detected.

\subsection{The Pan-STARRS 1 Survey}
The Pan-STARRS 1 telescope has surveyed large amounts of sky multiple 
times to moderate depths at declinations greater than -30~degrees.
We consider the analysis of the data in two stages.

 {\it The Pan-STARRS Survey for Transients. }
Like the earlier, CRTS, the Pan-STARRS Survey for Transients (PST)
\citep{2014ATel.5850....1S}
quickly disseminates detections of transients sources detected
in the sky. We have performed a similar analysis on the 
reported PST data, searching for viable Keplerian orbits.
As of 15 May 2016, no series of transients can be found which fit an outer solar 
system body on any bound Keplerian orbit. 
Typical transient depths reached are $g=21.0$, and,
based on the collection
of reported transients, the survey appears to efficiently cover
the sky north of -30 declination and at galactic latitudes
greater than about 10~degrees.
This survey rules out much of the sky within about 45~degrees
of the predicted perihelion point, with the exception of
the region near the galactic plane.

 {\it The Pan-STARRS Outer Solar System Key Project. } A survey
for objects in the outer solar system was one of the initial 
goals of the Pan-STARRS survey. \citet{2015DPS....4721112H} have now
completed a preliminary analysis of the survey data and report 
no detections out to 600~AU. While detailed sensitivity studies
have yet to be completed, it is estimated that the survey
is complete to approximately $r\sim 22.5$, though limits in
the galactic plane are worse. 
An extended analysis is currently underway
which will have the same brightness limits, but 
will remove the artificial restriction to objects closer
than 600~AU (Holman, private communication). 
If these sensitivity estimates are correct, the
Pan-STARRS 1 moving object survey has or will rule out a
substantial fraction of the non-aphelion sky.

\subsection{The Dark Energy Survey}
The Dark Energy Survey (DES) is performing the 
largest deep southern hemisphere survey to date.
Some of the DES 
region covers the orbital path of Planet Nine (indeed one of the
7 cluster objects, 2013 RF98, was detected in the DES). While cadences are
not designed for ease of outer solar system detection, it is clear that the
data will be sensitive to Planet Nine if it is in the survey
area. The DES team estimates a Planet Nine detection limit of $r\sim23.8$. 
(Gerdes, private communication). The survey should be completed in 2018.

\subsection{Additional surveys}
Additional surveys covering wide areas of the sky have been performed,
but in all cases they are insensitive to the slow expected speeds
of Planet Nine, they have too low of a survey efficiency
to consider the region effectively surveyed,
or they cover little or none of the region of the predicted orbital
path. The large community surveys which are concentrating
on specific areas of the sky, such as the VISTA surveys in the southern
hemisphere and the Subaru Hyper-SuprimeCam survey along the
celestial equator unfortunately do not overlap with the required
search region.

In the future, the Large-Scale Synoptic Telescope is expected to survey
much of the sky observable from its Chilean site to a single-visit
depth of approximately $r\sim 24.5$ magnitude. 
The current survey strategy does
not include visits to fields as far north as those at the extremes
of the predicted Planet Nine orbital path, but if Planet Nine has not
yet been found by the expected start of the LSST survey operations in
2023, a simple extension could quickly rule out nearly all but the
faintest and most distant Planet Nine predictions.

At its most distant predicted locations, Planet Nine is 
faint and in the northern hemisphere. Subaru Hyper-Suprime Cam will be the
instrument of choice for detecting the planet at these locations. 
We began a survey of these regions in the fall of 2015 and 
will attempt to cover all of this part of the predicted orbital path.

\subsection{Additional constraints:}
\citet{2016A&A...587L...8F}
perform full fits to the locations of all planets and
nearly 300 asteroids observed from ancient times to the present
with and without a 10~M$_e$ Planet Nine at various positions along an orbit with
$a_9=700$~AU and $e_9$=0.6, consistent with the nominal 
orbit suggested in \citet{2016AJ....151...22B}.
They find that the strongest constraint on the existence of Planet Nine
comes from the very precise measurements of the distance to Saturn
as measured by the Cassini spacecraft over 
the past decade.
With no Planet Nine, their best fit to the position
of Saturn has Earth-Saturn distance residuals which roughly follow a sinusoid with a
12-year period and a full amplitude of $\sim$70 m.  Previous
fits to the ephemeris of Saturn using the identical data, however,
found smaller residuals and no evidence of systematic variation
\citep{2014PhRvD..89j2002H}, while even more recent fits, leading to the creation of
the DE435 JPL planetary ephemeris, put even tighter constraints on 
any remaining residuals. It thus remains unclear whether any residual
is present in the Earth-Saturn distance.
We remain agnostic about the existence of
residuals in the distance to Saturn, but instead assume that the signal
apparently detected by \citet{2016A&A...587L...8F} is at or near the
level of the systematic errors in this type of analysis and that 
larger signals from Plant Nine could be detected. 
At its the nearest possible locations, 
in particular, Planet Nine would cause
such a large effect on the Earth-Saturn distance
that  the \citet{2016A&A...587L...8F}
analysis and an extension by \citet{2016arXiv160403180H} (which
assumes the same residuals)
can rule out any of the Planet Nine solutions near perihelion
over the range of Right Ascension of RA$_9>255$, RA$_9<2$ deg.

A recent paper \citep{2016arXiv160302196M} makes no attempt
to explain spatial alignments, but instead attempts to simplistically 
look for mean-motion commensurabilities in the distant KBOs, 
in hopes of being able to constrain
both $a_9$ and the location of Planet Nine within its orbit. 
Specifically they assume that the four most distant KBOs are in
N:1 and N:2 resonances, 
and examine the implications for Planet Nine. 
Such an approach could, in principle, work in the circular
restricted three-body problem, but, as shown in \citep{2016AJ....151...22B},
highly elliptical orbits are required to explain the spatial 
confinement of the orbits, and 
no specific resonances dominate the disturbing function in this
elliptical problem. Indeed, no particular preference for type of 
critical angle or even resonance order can be 
identified in the dynamical simulations shown here. 
Rather,
the crossing orbits evolve chaotically but
maintain long term stability by residing on 
a interconnected web of phase-protected mean motion resonances.
The assumption of simple low order resonance
is thus unlikely to be justified.  Not surprisingly, 
the Planet Nine orbits produced by these assumptions
do not produce the spatial confinements
of the KBOs that are observed.
Thus, it appears that no useful 
constraint on the orbit or position can be drawn from this method.

\subsection{Joint constraints}
In Figure 10 we show the regions of the potential orbits
of Planet Nine that have been ruled out 
by the above constraints (or, in the case of the ongoing DES and final
Pan-STARRS analysis,
where the planet might still be found by these surveys). 
The existence of Planet Nine can
be ruled out over about two thirds of its orbit. 
The vast majority of the orbital region in which Planet Nine 
could be located is beyond about 700~AU
and within about 60~degrees of its aphelion 
position.
For the eccentric orbits
considered here, Planet Nine spends greater than half of its time
at these distances, so finding it currently at these locations
near aphelion would be expected.
At its most distant allowed location and with a Neptune-like albedo,
a 20 R$_e$ Planet Nine is approximately $V=25$. While faint, such
an object would be well within the limits of 10-m class telescopes.

\section{Conclusions}
The existence of a distant massive perturber in the outer solar system --
Planet Nine -- explains several hitherto unconnected observations
about the outer solar system, including the orbital alignment of the
most distant Kuiper belt objects, the existence and alignment of high perihelion 
objects like Sedna, and the presence of perpendicular high semimajor axis
Centaurs. These specific observations have been compared to suites of 
numerical integrations in order to constrain possible parameters of Planet Nine. The current constraints must be considered preliminary: our orbital 
simulations needed to cover substantial regions of potential phase space, 
and so were, of necessity, sparsely populated. At present, the
statistical reliability
 of our constraints are limited as much by the limited
survey nature of
the simulations as by the small number of observed objects themselves.
Continued simulation could substantial narrow the potential search area
required. In addition, continued simulation is required in order to
understand one effect not captured in the current models: the apparent
alignment of argument of perihelion of the 16 KBOs with the largest semimajor
axes \citep{2014Natur.507..471T}. Some of this apparent alignment may come from
yet unmodeled observational biases related to the close proximity 
of the perihelion positions of the most distant objects to the galactic plane,
while some may be a true as-yet-unmodeled dynamical effect.

As important as continued simulation, continued detection of distant
solar system objects is the key to refining the orbital parameters of 
Planet Nine. Each addition Kuiper belt object (or Centaur) with $a>100$~AU 
tightens the observational constraints on the location of Planet Nine (or,
alternatively, if significant numbers of objects are found outside of
the expected cluster location, the objects can refute the presence of 
a Planet Nine). 

Interestingly, the detection of more
high semimajor axis perpendicular objects
(whether Centaurs or Kuiper belt objects) has the possibility of 
placing the strongest constraints in the near term. While we have
currently only used the {\it existence} of these objects as a constraint, their
perihelion locations and values of $\omega$ change strongly with
$\omega_9$ and $i_9$ and so can be used to better refine these
estimates. Though there are only currently 5 known of these objects, 
they are being discovered at a faster rate than the distant Kuiper
belt objects, so we have hope of more discoveries soon. As with
the distant Kuiper belt objects, of course, detection of
these objects also has the strong possibility of entirely 
ruling out the existence of Planet Nine if they are not found with
perihelia in the locations predicted by the hypothesis.

\clearpage

\clearpage

\end{document}